\documentclass[9pt,fleqn]{article}
\usepackage{amsmath}
\usepackage{amssymb}
\usepackage{amsfonts}
\usepackage{float}
\usepackage{epsfig}
\usepackage{cite}

\newenvironment{proof}[1][Proof]{\noindent\textbf{#1.} }{\ \rule{0.5em}{0.5em}}
\begin{document}

\title{Effects of heterogeneity on cancer: a game theory perspective\thanks{
Laruelle acknowledges financial support from the Spanish Ministry of Science
and Innovation under funding PID2019-106146-I00 and from the Basque
Government (Research Group IT1367-19); Rocha acknowledges financial support
from the Brazilian Ministry of Science, Technology and Innovations (CNPq
funding 307437/2019-1); Inarra acknowledges financial support from the
Spanish Ministry of Science and Innovation (project PID2019-107539GB-I00)
and from the Basque Government (project IT1367-19). }}
\author{{\Large A}nnick {\Large L}aruelle\thanks{%
Department of Economic Analysis (ANEKO), University of the Basque Country
(UPV/EHU); Avenida Lehendakari Aguirre, 83, E-48015 Bilbao, Spain;
annick.laruelle@ehu.eus.}\thanks{%
IKERBASQUE, Basque Foundation of Science, 48011, Bilbao, Spain.}, {\Large A}%
ndr\'{e}\ {\Large R}ocha\thanks{%
Department of Industrial Engineering, Pontifical Catholic University of Rio
de Janeiro, Rua Marqu\^{e}s de S\~{a}o Vicente 225, G\'{a}vea, CEP22451-900,
Rio de Janeiro, RJ, Brazil.}, {\Large C}laudia {\Large M}anini\thanks{%
Department of Pathology, San Giovanni Bosco Hospital, 10154 Turin, Italy.}%
\thanks{%
Department of Sciences of Public Health and Pediatrics, University of Turin,
10124 Turin, Italy.}, {\Large J}os\'{e} {\Large I}\ {\Large L}\'{o}pez%
\thanks{%
Department of Pathology, Cruces University Hospital, 48903 Barakaldo, Spain.}%
\thanks{%
Biomarkers in Cancer Group, Biocruces-Bizkaia Research Institute, 48903
Barakaldo, Spain.}, and {\Large E}lena\ {\Large I}narra\thanks{%
Department of Economic Analysis (ANEKO) and Public Economic Institute,
University of the Basque Country (UPV/EHU), Avenida Lehendakari Aguirre, 83,
E-48015 Bilbao, Spain.} }
\date{}
\maketitle

\begin{abstract}
In this study, we explore interactions between cancer cells by using the
hawk-dove game. We analyze the heterogeneity of tumors by considering games
with populations composed of 2 or 3 types of cells. We determine what
strategies are evolutionarily stable in the 2-type and 3-type population
games and the corresponding expected payoffs. Our results show that the
payoff of the best-off cell in the 2-type population game is higher than
that of the best-off cell in the 3-type population game. Translating these
mathematical findings to the field of oncology suggests that a
branching-type tumor pursues a less aggressive course than a punctuated-type
one. Some histological and genomic data of clear cell renal cell carcinomas
are consistent with these results.

 \vspace{0.5cm}

\textbf{Significance}

\noindent This study underlines the importance of identifying intratumor
heterogeneity in routine practice. It suggests that a tumor with high
intratumor heterogeneity is less aggressive than one with low intratumor
heterogeneity; that is, therapeutic strategies that preserve heterogeneity
may be promising as they may slow down cancer growth.
\end{abstract}

\baselineskip=0.6cm \oddsidemargin=0.5cm \evensidemargin=1.5cm \topmargin
=0cm \vsize=22cm \textheight=21cm \vfill\newpage

Intratumor heterogeneity (ITH)\ is a constant in malignant neoplasms and the
factor responsible for most therapeutic failures in clinical practice.
Recent advances provided by massive genomic sequencing are showing the real
scale and variability of spatial and temporal evolutionary patterns in many
tumors \cite{Dentro21,Tong18,Thomas20}. Tumor evolution was initially
considered to be a stochastic process, but some deterministic pathways
correlating with prognosis have been identified recently \cite{Turajlic18}.
Along this line of research, several models have been described (linear,
branching, neutral, and punctuated) \cite{Davis17}, reflecting different
evolutionary patterns that promote spatial and temporal clonal and
sub-clonal ITH. This paper compares branching and punctuated patterns of
tumor evolution.

In oncology, fitness is a direct measure of the capacity of tumor cells to
grow and metastasize, i.e. \ of their degree of biological aggressiveness
and ability to kill the patient. Cells with better adapted traits in the
struggle for existence grow faster in a tumor. This evidence supports the
use of game theory to analyze cell interactions, cell fitness and tumor
evolution \cite{Axelrod18, Dingli09, Archetti19, Wolf21}. In particular the
hawk-dove game has been used to explore competition between cells for an
external resource \cite{McEvoy09} when obtaining that resource may be costly.

Hawk-dove games are defined by two parameters: the size of the resource ($v$%
) and the size of the cost ($c$). In tumors the resource consists of a
varied spectrum of diffusible factors released into the medium by tumor
and/or non-tumor cells. An example is the fibroblast activation protein-$%
\alpha $ produced and released by a sub-type of cancer-associated
fibroblasts \cite{Errarte20}. Resource necessities increase with the
histological grade of the tumor. Several studies show that tumor cells
develop a feedback mechanism to increase resource production \cite%
{Archetti21}. It can be assumed that the higher the grade the greater the
resources. The cost is the energy required to obtain the resource. The
Atkinson level enables the energy spent by any cell, normal or neoplastic,
to be measured. It takes into account the relative cytoplasmic
concentrations of ATP, ADP, and AMP \cite{DelaFuente14}. The resource and
the cost are linked to the grade of the tumor, and are assumed to be
constant for a given grade.

The novelty of this study is that we analyze the heterogeneity of tumors by
using a heterogeneous hawk-dove game \cite{Inarra12}. That is, instead of
considering a homogeneous population we consider heterogeneous populations
composed of two or three types of cell. In our model, ITH is considered as
low when the tumor has two different histologies. Punctuated-type tumors are
represented here by 2-type population games. When the tumor has three or
more identifiable histologies ITH is high. These are tumors with a branching
pattern of evolution, which are represented here by 3-type population games.
A hypothetical homogeneous cancer would be represented by 1-type population,
but this possibility is unrealistic in medical practice.

Examples of how cancer cells behave differently depending on the recognition
of their respective cellular contexts are available in the literature \cite%
{Maruyama17}. We thus assume that cells detect the type of their opponents
but do not recognize their own type. In other words, cells are genetically
programmed to play strategies which are not conditional on their own type
but on that of their opponent.

The solution concept applied to solve these games is the evolutionarily
stable strategy (ESS) \cite{Maynard82,Maynard73,Maynard76}. An ESS is a
strategy that maximizes the expected payoff of a player when its opponent
chooses the same strategy, i.e. it is a symmetric Nash equilibrium. An ESS
also guarantees that no mutant strategy can invade the population.

We determine the ESS for the 2-type and 3-type population games. In the ESS
different cells of different types obtain different expected payoffs. The
best-off cells are those with the highest expected payoffs: These divide at
a higher rate than any other, increasing their number in the tumor. They are
the most malignant and the ones that determine the aggressiveness of the
tumor. Thus, we compare the expected payoffs of the best-off cells in the
2-type and 3-type population games, and find the conditions under which the
best-off cells in the 2-type population game obtain a larger expected payoff
than the best-off cells in the 3-type population game. When these conditions
are not met the payoffs obtained in the 2-type and 3-type population games
are equal.

\section*{The hawk-dove game in tumors}

We consider a tumor formed by a population of $n$ cells. Encounters between
cells are bilateral and in each encounter a cell can behave aggressively,
like a hawk, or passively, like a dove, in order to acquire a resource $v$.
If a cell is aggressive and its opponent is passive, the aggressive cell
obtains the resource and the other gets nothing. If both cells are
aggressive there is a fight and the winner gets the resource while the loser
bears a cost $c>v$. Assuming that they both have the same probability of
winning, the expected payoff for each cell is $(v-c)/2$. If both cells are
passive, one withdraws and gets nothing while the other takes the resource.
Assuming they both have the same probability of withdrawing, the expected
payoff for each cell is $v/2$. These contingencies are summarized in the
following payoff matrix.\vspace{-0.95cm}

\begin{center}
\begin{equation*}
\begin{tabular}{c|cc}
& hawk & dove \\ \hline
hawk & $\frac{v-c}{2}$ & $v$ \\ 
dove & $0$ & $\frac{v}{2}$%
\end{tabular}%
\end{equation*}
\end{center}

A strategy is given by $\alpha $, the probability of a cell being aggressive
on meeting another cell. A cell can choose to play either hawk ($\alpha =1$)
or dove ($\alpha =0$) or a mixed strategy ($0<\alpha <1$). The expected
payoff of a cell that plays $\alpha $ when its opponent plays $\beta $ is
given by $u(\alpha ,\beta )$: 
\begin{equation}
u(\alpha ,\beta )=\frac{v}{2}(1-\beta )+\frac{c}{2}\left( \frac{v}{c}-\beta
\right) \alpha .  \label{U1}
\end{equation}%
Strategy $v/c$ is the only ESS \cite{Maynard82} when the population is
homogeneous, i.e. composed of a single type of player.

\textbf{Heterogeneous population.} Tumors are usually composed of different
types of cells. We consider two levels of ITH: Low and high. In our model a
tumor with low ITH is composed of two types of cells: $A$-cells and $B$%
-cells. A tumor with high ITH is composed of three types of cells: $A$%
-cells, $B$-cells and $E$-cells.

As shown below, heterogeneity induces a differentiation in terms of payoffs
in the ESS. We set the following hierarchy: $A$-cells are the best-off
cells, i.e. those with the strictly largest expected payoff. In the 3-type
population $E$-cells are the worst-off cells, i.e. those with the strictly
smallest expected payoff. The fittest cells are those that determine tumor
aggressiveness. Thus, we compare the expected payoff of the best-off $A$%
-cells in tumors with low and high ITH.

We seek to analyze the effect of ITH on the fitness of tumors. We thus
compare cells with identical capacities in terms of both the level of
resources to which they are exposed and their energy cost in fighting for
those resource. A priori types do not confer any advantage: The same payoff
matrix is played in every encounter. Also, as mentioned in the introduction,
cells are sensitive to their environment: They recognize the type of their
opponent and can adapt their behavior depending on the opponent that they
meet. They do not behave according to their own type.

\textbf{Tumors with low ITH.} Consider a heterogeneous tumor formed by $A$%
-cells and $B$-cells. A 2-type population game is denoted by $\Gamma
_{2}(v,c,x_{A})$, as it can be defined by parameters $v$, $c$, and the
proportion of $A$-cells, $x_{A}$. Cells cannot adopt a different strategy
according to their own type but are able to choose a distinct probability of
playing aggressively when facing any type of opponent. Thus, a strategy is a
pair ($\alpha _{A},\alpha _{B}$), where $\alpha _{I}$ denotes the
probability of behaving aggressively when meeting an $I$-cell ($I=A,B$). In
other words, $\alpha _{I}$ indicates the level of aggression received by an $%
I$-cell. Game $\Gamma _{2}(v,c,x_{A})$ is analyzed in \cite{Inarra12}. It is
shown that no strategy with $\alpha _{A}=\alpha _{B}$ is evolutionarily
stable and in the ESS cells of one type receive less aggression and a larger
expected payoff than cells of the other type.

We focus on the ESS where the A-cells have the largest expected payoff. This
ESS, which depends on the proportion of $A$-cells, is given by: 
\begin{equation*}
\left\{ 
\begin{array}{ll}
\left( 0,\tfrac{n-1}{n-nx_{A}-1}\tfrac{v}{c}\right) & \text{if }x_{A}<\bar{x}%
_{A} \\ 
(0,1) & \text{if }\bar{x}_{A}<x_{A}<\bar{x}_{A}+\frac{1}{n} \\ 
\left( \frac{(n-1)v/c-n+nx_{A}}{nx_{A}-1},1\right) & \text{if }x_{A}>\bar{x}%
_{A}+\frac{1}{n},%
\end{array}%
\right.
\end{equation*}%
where $\bar{x}_{A}=\left( 1-\frac{v}{c}\right) \left( 1-\frac{1}{n}\right) $.

\begin{figure}[H]
\centering
\begin{tabular}{c}
\epsfig{file=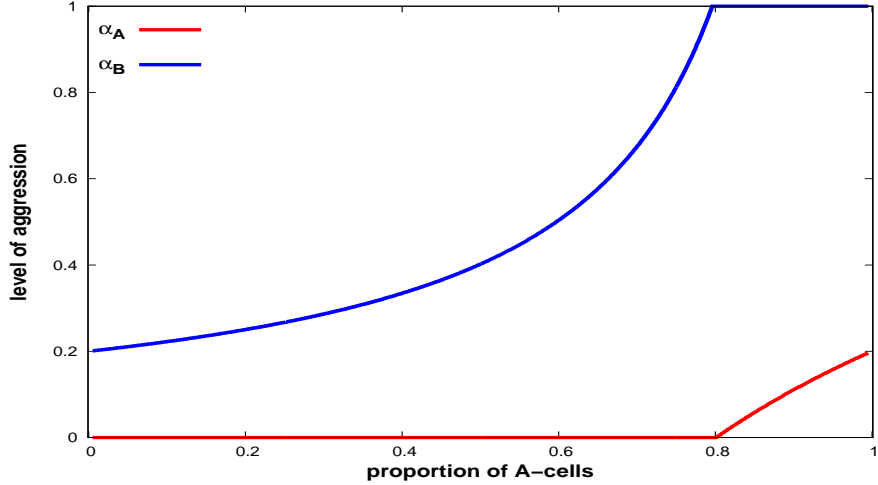,height=6.5cm,width=12cm,angle=0}%
\end{tabular}
\vspace{-.15cm}
\caption{Aggression received at the ESS by $A$-cells ($\protect\alpha _{A}$)
and $B$-cells ($\protect\alpha _{B}$) in a 2-type population game as a
function of $x_{A}$. Parameters: $v=10$, $c=50$, $n=201$ and $\bar{x}%
_{A}=0.796$.}
\label{fig:1}
\end{figure}

In the ESS the best-off $A$-cells receive less aggression than $B$-cells ($%
\alpha _{A}<\alpha _{B}$) do. The level of aggression toward the latter
increases when the proportion of $A$-cells increases, until full aggression (%
$\alpha _{B}=1$) is reached (for $x_{A}>\bar{x}_{A}$). Also, if the
proportion of $A$-cells is below a threshold ($x_{A}<\bar{x}_{A}+1/n$), $A$%
-cells suffer no aggression ($\alpha _{A}=0$) and aggression is concentrated
only on $B$-cells. Above that threshold, however, $A$-cells do receive some
aggression. Figure \ref{fig:1} plots the level of aggression received in the
ESS by $A$-cells ($\alpha _{A}$) and $B$-cells ($\alpha _{B}$) as a function
of the proportion of $A$-cells, $x_{A}$. It shows that the aggression toward 
$A$-cells and $B$-cells increases as $x_{A}$ gets larger.

Since only the expected payoff of the best-off cells is of interest in this
context, we focus on the expected payoff in the ESS of the best-off $A$%
-cells, denoted by $U_{A}^{\ast \ast }(x_{A})$. That is given by: 
\begin{equation}
U_{A}^{\ast \ast }(x_{A})=\left\{ 
\begin{array}{ll}
\tfrac{v}{2}(1-\tfrac{v}{c})+\tfrac{v^{2}}{2c}\tfrac{2n(1-x_{A})-1}{%
n-nx_{A}-1} & \text{if }x_{A}<\bar{x}_{A} \\ 
\tfrac{v}{2}(1-\tfrac{v}{c})+\tfrac{v}{2c}\tfrac{v(n-1)+cn(1-x_{A})}{n-1} & 
\text{if }\bar{x}_{A}\leq x_{A}\leq \bar{x}_{A}+\frac{1}{n} \\ 
\tfrac{v}{2}(1-\tfrac{v}{c})+\tfrac{c-v}{c}\tfrac{vn(1-x_{A})}{nx_{A}-1} & 
\text{if }x_{A}>\bar{x}_{A}+\frac{1}{n}.%
\end{array}%
\right.  \label{U2}
\end{equation}%
If the proportion of $A$-cells is below a threshold ($x_{A}<\bar{x}_{A}$),
in the ESS the larger $x_{A}$ is, the larger the expected payoff of the $A$%
-cells (see SI-Appendix A). Indeed, $A$-cells receive no aggression while $B$%
-cells receive more aggression as $x_{A}$ increases. In consequence, an $A$%
-cell more frequently obtains resource $v$, thus increasing its expected
payoff. The expected payoff of an $A$-cell peaks when $x_{A}=\bar{x}_{A}$.
Thereafter, the expected payoff of an $A$-cell starts decreasing. The
explanation is that $A$-cells start fighting among themselves for resource $%
v $.

\textbf{Tumors with high ITH.} Consider a heterogeneous tumor formed by $A$%
-cells (in proportion $x_{A}$), $E$-cells (in proportion $x_{E}$) and $B$%
-cells (in proportion $x_{B}=1-x_{A}-x_{E}$). A 3-type population game is
denoted by $\Gamma _{3}(v,c,x_{A},x_{E})$, as it can be defined by
parameters $v$, $c$, and the proportions of $A$-cells and $E$-cells.

A strategy is denoted by ($\alpha _{A},\alpha _{B},\alpha _{E}$) where $%
\alpha _{I}$ is the probability of behaving aggressively when facing an $I$%
-cell ($I=A,B,E$). If an $I$-cell plays ($\alpha _{A},\alpha _{B},\alpha
_{E} $) against an opponent playing ($\beta _{A},\beta _{B},\beta _{E}$) its
expected payoff is the sum of the probability of meeting a cell of its own
type, i.e., $(nx_{I}-1)/(n-1)$, multiplied by $u(\alpha _{I},\beta _{I})$,
and the probability of meeting a cell of each other cell type, i.e., $%
(nx_{J})/(n-1)$, multiplied by $u(\alpha _{J},\beta _{I})$. Denoting the
expected payoff of an $I$-cell by $U_{I}$, the following is obtained: 
\begin{eqnarray}
U_{A} &=&\frac{nx_{A}-1}{n-1}u(\alpha _{A},\beta _{A})+\frac{n(1-x_{A}-x_{E})%
}{n-1}u(\alpha _{B},\beta _{A})+\frac{nx_{E}}{n-1}u(\alpha _{E},\beta _{A}) 
\notag \\
U_{B} &=&\frac{nx_{A}}{n-1}u(\alpha _{A},\beta _{B})+\frac{n(1-x_{A}-x_{E})-1%
}{n-1}u(\alpha _{B},\beta _{B})+\frac{nx_{E}}{n-1}u(\alpha _{E},\beta _{B}) 
\notag \\
U_{E} &=&\frac{nx_{A}}{n-1}u(\alpha _{A},\beta _{E})+\frac{n(1-x_{A}-x_{E})}{%
n-1}u(\alpha _{B},\beta _{E})+\frac{nx_{E}-1}{n-1}u(\alpha _{E},\beta _{E}).
\label{U3I}
\end{eqnarray}%
Observe that whenever cells choose the same strategy, which in equilibrium
is always the case, the larger the payoff is the smaller the aggression that
a cell type receives is. That is, $U_{I}>U_{J}$ is equivalent to $\alpha
_{I}<\alpha _{J}$ (for $I,J\in \{{A,B,E\}}$).

Given that a cell does not know its own type, what can be maximized is the
expected payoff of a generic cell, denoted by $U$, given by the sum of the
probability of being $I$ multiplied by the expected payoff of an $I$-cell: 
\begin{equation}
U=x_{A}U_{A}+(1-x_{A}-x_{E})U_{B}+x_{E}U_{E}.  \label{U}
\end{equation}%
To be consistent with the hierarchy of cell types established above, we
focus on the ESS where the three different types receive different expected
payoffs such that $U_{A}>U_{B}>U_{E}$. The following proposition (See
SI-Appendix B) indicates that the ESS depends on the proportions of the $A$%
-cells and $E$-cells.

\begin{figure}[H]
\centering
\begin{tabular}{c}
\epsfig{file=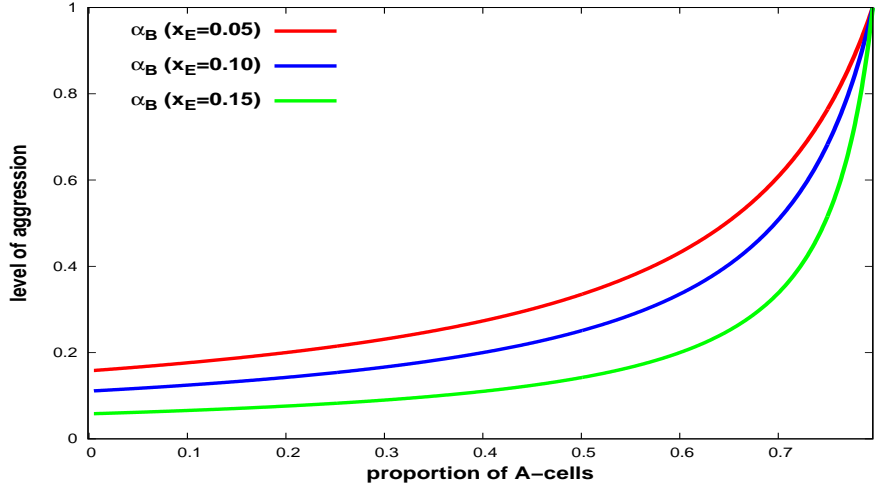,height=6.5cm,width=12cm,angle=0}%
\end{tabular}
\vspace{-.15cm}
\caption{Aggression received at the ESS by $B$-cells ($\protect\alpha _{B}$)
in a 3-type population game as a function of $x_{A}$ for different values of 
$x_{E}$. Parameters: $v=10$, $c=50$, $n=201$, $\bar{x}_{A}=0.796$ and $\bar{x%
}_{E}=0.199$.}
\label{fig:2}
\end{figure}

\textbf{Proposition 1:} Let $\Gamma _{3}(v,c,x_{A},x_{E})$ be a 3-type
population game. Thus, there exists an ESS with $U_{A}>U_{B}>U_{E}$ if and
only if $x_{A}<\bar{x}_{A}$ and $x_{E}<\bar{x}_{E}$, where $\bar{x}_{E}=%
\frac{v}{c}\left( 1-\frac{1}{n}\right) $. The ESS is given by: 
\begin{equation}
\left( 0,\frac{(n-1)v/c-nx_{E}}{n-nx_{A}-nx_{E}-1},1\right) .  \label{ESS3}
\end{equation}%
As expected, $A$-cells receive less aggression than $B$-cells, which in turn
receive less aggression than $E$-cells ($\alpha _{A}<\alpha _{B}<\alpha _{E}$%
). The best-off cells do not receive any aggression, $B$-cells receive some
aggression and $E$-cells receive full aggression. Figure \ref{fig:2} plots
the aggression suffered by $B$-cells in the ESS for different proportions of 
$A$-cells and $E$-cells, showing that: (i) the larger $x_{A}$, the greater
the aggression suffered by $B$-cells; and (ii) the larger $x_{E}$, the
lesser the aggression suffered by $B$-cells (see SI-Appendix C).

Since only the expected payoff of the best-off $A$-cells is of interest in
this context, we present the expected payoff in the ESS of those cells,
denoted by $U_{A}^{\ast \ast \ast }(x_{A},x_{E})$. This is given by: 
\begin{equation}
U_{A}^{\ast \ast \ast }(x_{A},x_{E})=\frac{v}{2}\left( 1-\frac{v}{c}\right) +%
\frac{v^{2}}{2c}\frac{2n(1-x_{A}-x_{E})-1}{n-nx_{A}-nx_{E}-1}-\frac{v}{2}%
\frac{nx_{E}}{\left( n-nx_{A}-nx_{E}-1\right) \left( n-1\right) }.
\label{U3}
\end{equation}%
The larger $x_{A}$ is, the larger the expected payoff of each $A$-cell is
(See SI-Appendix D). Indeed, the aggression towards $B$-cells increases
while $A$-cells receive no aggression (and $E$-cells receive full
aggression). As a result, an $A$-cell obtains resource $v$ more often and
its expected payoff therefore increases. By contrast, the larger $x_{E}$ is,
the smaller the expected payoff of each $A$-cell is. As $x_{E}$ increases
two opposite effects are at play: On the one hand there are more $E$-cells,
which receive full aggression, increasing overall aggression levels, which
is beneficial for $A$-cells. On the other hand, as $x_{E}$ increases the
aggression received by $B$-cells decreases, which reduces overall
aggression, which is detrimental for $A$-cells. This negative effect turns
out to outweigh the positive effect, so the expected payoff of each $A$-cell
decreases with $x_{E}$. 

An important point to highlight here is that there is no ESS with $%
U_{A}>U_{B}>U_{E}$ when the proportion of $A$-cells or $E$-cells exceed
thresholds $\bar{x}_{A}$ or $\bar{x}_{E}$. Having $x_{A}>\bar{x}_{A}$ or $%
x_{E}>\bar{x}_{E}$ does not prevent the existence of ESS. In these cases two
cell types receive an identical expected payoff: Either $U_{A}>U_{B}=U_{E}$
or $U_{A}=U_{B}>U_{E}$, thus reproducing a 2-type population game. This can
be illustrated for the case $U_{A}>U_{B}=U_{E}$.

\textbf{Proposition 2:} Let $\Gamma _{3}(v,c,x_{A},x_{E})$ be a 3-type
population game. There exists an ESS with $U_{A}>U_{B}=U_{E}$ for $x_{A}>%
\bar{x}_{A}$, given by 
\begin{equation*}
\left\{ 
\begin{array}{ll}
(0,1,1) & \text{if }\bar{x}_{A}<x_{A}<\bar{x}_{A}+1/n \\ 
\left( \frac{(n-1)v/c-n(1-x_{A})}{nx_{A}-1},1,1\right) & \text{if }x_{A}>%
\bar{x}_{A}+1/n.%
\end{array}%
\right.
\end{equation*}%
In this ESS, the $E$-cells are not distinguished from the $B$-cells ($\alpha
_{B}=\alpha _{E}$). The level of aggression received and the expected
payoffs are identical to those obtained in the ESS in the 2-type population
game.

\textbf{Comparison of heterogeneous tumors.} The main question here concerns
the comparison of the expected payoff of the best-off $A$-cells in the
2-type and 3-type population games. Clearly, the comparison is pertinent for
equal proportions of $A$-cells in the two tumors under analysis. The
following proposition (See SI-Appendix E) indicates when the expected payoff
of an $A$-cell in ESS is strictly larger in the 2-type population game than
in the 3-type population game. 

\textbf{Proposition 3:} Let $y_{A}$ be a proportion of A-cells and let $%
\Gamma _{2}(v,c,y_{A})$ and $\Gamma _{3}(v,c,y_{A},x_{E})$ with $y_{a}<\bar{x%
}_{A}$ and $x_{E}<\bar{x}_{E}$. Then $U_{A}^{\ast \ast }(y_{A})>U_{A}^{\ast
\ast \ast }(y_{A},x_{E})$.

\begin{figure}[H]
	\centering
	\begin{tabular}{cc}
		\epsfig{file=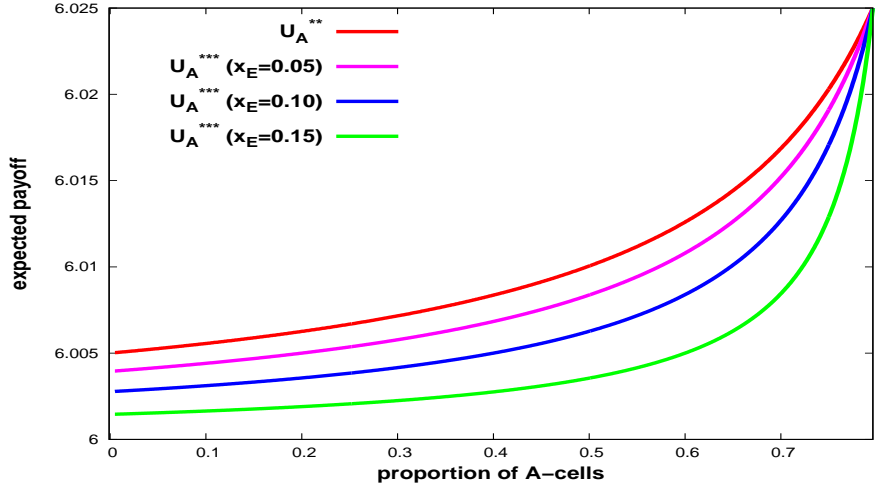,height=6.5cm,width=12cm,angle=0} & 
	\end{tabular}
	\vspace{-.15cm}
	\caption{Expected payoff of the best-off $A$-cells at the ESS in a 2-type
		population game ($U_{A}^{\ast \ast }$) and 3-type population game ($%
		U_{A}^{\ast \ast \ast }$ for different values of $x_{E}$) as functions of $%
		x_{A}$. Parameters: $v=10$, $c=50$, $n=201$, $\bar{x}_{A}=0.796$ and $\bar{x}%
		_{E}=0.199$.}
	\label{fig:3}
\end{figure}

This result is illustrated in Figure \ref{fig:3}. Recall that the ESS for $%
x_{A}<\bar{x}_{A}$ in $\Gamma _{2}(v,c,x_{A})$ is $\left( 0,\tfrac{n-1}{%
n-nx_{A}-1}\tfrac{v}{c}\right) $ while the ESS in $\Gamma
_{3}(v,c,x_{A},x_{E})$ is $\left( 0,\frac{(n-1)v/c-nx_{E}}{n-nx_{A}-nx_{E}-1}%
,1\right) $. In both games the best-off $A$-cells receive no aggression,
while $E$-cells receive full aggression in the 3-type population game. $B$%
-cells receive some aggression in both games. The aggression received by $B$%
-cells in the 2-type population game is greater than that which they receive
in the 3-type population game. This difference confers an advantage to $A$%
-cells in the former.
\textbf{Translation in the oncology context}. Proposition 1 indicates that
an ESS in a 3-type population game only exists when the proportions of the
best-off and worst-off cells do not exceed certain thresholds, $\bar{x}_{A}$
and $\bar{x}_{E}$. Therefore patients with tumors displaying a high ITH and
a large proportion of the highest grade cells are not expected to be found.
In other words, branching-type tumors should not display predominant
proportions of the highest grade cells.

According to Proposition 2, beyond the thresholds for the best-off and
worst-off cells, there is no ESS that distinguishes between the three types
of cells in the 3-type population games since only two types can be
distinguished. This means that once the proportion of the highest grade
cells becomes large the ITH decreases. A branching-type tumor may thus
become punctuated.

Proposition 3 shows that the best-off cell obtains a larger expected payoff
in 2-type population games than in 3-type population games (for equal
proportions of $A$-cells). This means that if two tumors with equal
proportion of the highest grade are compared the one with the higher ITH
should be less aggressive than the one with the lower ITH. In other words,
for a given grade a branching-type tumor pursues a less aggressive course
than a punctuated-type one.

\section*{Proof of concept}
The pathologists in the study retrospectively analyzed a
series of 20 clear cell renal cell carcinomas (CCRCC) from archives of the
Department of Anatomic Pathology at Cruces University Hospital (Barakaldo,
Spain). Table \ref{tab1} gives the main data of the series: Basic data on
the patient (age and sex), number of regions analyzed, characteristics of
the tumors (grade, heterogeneity), follow-up information, and the clinical
situation of the patient. 
\begin{table}[H]
\centering
\begin{tabular}{|c|c|c|c|c|c|c|c|}
\hline
Case & sex & age & regions & highest G (\%) & heterogeneity & follow-up & 
outcome \\ \hline
1 & M & 73 & 56 & 3 (10) & low & 67 & AWD \\ 
2 & F & 50 & 60 & 4 (5) & high & 65 & AwoD \\ 
3 & M & 64 & 60 & 4 (80) & low & 69 & AwoD \\ 
4 & M & 70 & 40 & 2 (100) & no & 68 & AwoD \\ 
5 & M & 46 & 66 & 3 (60) & low & 65 & AWD \\ 
6 & M & 42 & 40 & 1 (100) & no & 63 & AwoD \\ 
7 & M & 70 & 64 & 4 (5) & high & 64 & AwoD \\ 
8 & F & 41 & 55 & 4 (10) & high & 67 & AwoD \\ 
9 & M & 53 & 48 & 4 (70) & low & 50 & DOD \\ 
10 & M & 56 & 54 & 4 (90) & low & 61 & DOD \\ 
11 & M & 66 & 81 & 4 (90) & low & 11 & DOD \\ 
12 & M & 67 & 48 & 4 (70) & low & 16 & DOD \\ 
13 & M & 71 & 77 & 4 (80) & low & 49 & DOD \\ 
14 & F & 68 & 59 & 4 (70) & low & 5 & DOD \\ 
15 & M & 77 & 56 & 3 (5) & high & 68 & AwoD \\ 
16 & M & 70 & 54 & 3 (10) & low & 62 & AWD \\ 
17 & F & 51 & 52 & 3 (5) & high & 61 & AwoD \\ 
18 & M & 82 & 69 & 4 (50) & low & 45 & DOD \\ 
19 & M & 62 & 57 & 4 (90) & low & 64 & AWD \\ 
20 & F & 61 & 74 & 3 (20) & high & 62 & AwoD \\ \hline
\end{tabular}%
\caption{regions: number of regions sampled and analyzed; highest G (\%):
highest histological grade detected across all the regions and its
percentage; heterogeneity: grade heterogeneity, low (two different grades)
vs. high (three or more different grades); follow-up: follow-up in months;
outcome: outcome at the last contact; AWD: alive with disease; AwoD: alive
without disease; DOD: died of disease.}
\label{tab1}
\end{table}

The histological grade of a tumor is one of several measures that define its
level of malignancy, i.e. biological aggressiveness. The ways in which
pathologists assign a grade to a tumor under the microscope vary widely from
one tumor entity to another. In general, the spectrum of grades in a tumor
oscillates between grades 1 and 4, where the higher the grade is, the more
aggressive the tumor is expected to be. The proportion of the highest
nuclear grade detected appears in the table. In our modeling, this
corresponds to the proportion of the best-off $A$-cells, $x_{A}$.

ITH measures the cellular variability detected in the tumor, a
characteristic which is routinely assessed in modern pathology. Two levels
are considered in the table: Low ITH is considered to exist when no more
than two different grades are detected and high ITH when three or more
different grades are seen.

Of the 12 patients with tumors showing low ITH, seven died of disease, four
were alive with disease, and one was alive without the disease at last
contact. All patients with tumors showing high ITH were alive without the
disease at last contact. Comparative survival rates show that the difference
between low and high ITH is statistically significant (log-rank $p<0.001$).
Note that the comparison involves tumors with different grades and different
proportions. A higher grade and a large proportion of the highest grade are
generally associated with a worse clinical evolution of the patient. This
may explain the prognoses of patients with grade 4 tumors and high
proportions of the most malignant cells, all but two of whom (patients 3 and
19) died of the disease.

To check Proposition 3 we need to compare tumors of identical grades and
similar (if not identical) proportions of the highest grade. Five out of the
six grade 3 tumors have similar proportions of the highest grade ($x_{A}$
between $5\%$ and $20\%$). It can be observed that the three patients with
tumors that show high ITH (patients 15, 17, and 20) are alive without the
disease while the two with low ITH tumors are alive with the disease (1, and
16). This finding is consistent with Proposition 3, assuming that $\bar{x}%
_{A}>20\%$. Note that patient 20, with a highly heterogeneous tumor of
proportion $x_{A}=20\%$, is alive without the disease, while patients 1 and
16 with low heterogeneous tumors of proportion $x_{A}=10\%$, are alive with
the disease.

According to Proposition 1 patients with tumors displaying a high ITH and a
large proportion of the highest grade cells are not expected to be found. We
indeed observe that the six tumors with high ITH have small proportions of
the highest grade cells ($x_{A}<20\%$).

\section*{Discussion}

This paper looks at the hawk-dove game in heterogeneous populations. The
main assumption in this context is that cells are programmed to adopt a
specific behavior according to the behavior of cells with which they
interact. As a result, different cells are treated differently and obtain
different fitness levels. The difference in terms of fitness in a cell
depends only on the different behaviors adopted by the remaining cells, and
not on its own behavior.

The spectrum of cell variability in a tumor is an important characteristic
routinely assessed in modern pathology. Our study suggests that this
characteristic may be correlated with tumor aggressiveness. Here,
mathematical findings are supported by histological and genomic data. For
example, a basic histological study of 20 CCRCC exhaustively sampled and
followed clinically for at least 5 years confirms that aggressive forms of
CCRCC show low levels of ITH. Moreover, a genomic study of a series of 101
CCRCC has shown that tumors with low ITH and a punctuated-type evolutionary
pattern pursue an aggressive clinical course, with early and multiple
metastases, whereas tumors with high ITH and branching-type evolution show a
more attenuated clinical course, with late, solitary metastases \cite%
{Turajlic18}. Our study suggests also that branching-type tumors may, under
specific conditions, evolve into punctuated-type tumors. This evolution has
been detected in several types of malignant tumors \cite{Bao21}.

At a practical level, our results support the idea that it is appropriate to
preserve high ITH in tumors as a promising therapeutic strategy. Thus, a
therapy aimed at tumor containment should be more effective than
administrating the conventional maximum tolerable dose because the latter
strategy would force tumor cells to develop different clones able to survive
as soon as possible \cite{Gundem15}. As a result, a resistant-to-therapy
neoplasm develops. Even if the tumor presents a high ITH before therapy,
recurrence will result in a low ITH neoplasm. The maximum tolerable dose has
already been questioned using a game theory approach \cite{Archetti21}.

The strategy of tumor containment seeks to diversify the energy expenditure
of the cell by administrating drugs below the maximum tolerable dose \cite%
{Gundem15}. Tumor cells will thus distribute their energy across different
tasks, causing them to lose efficiency in all of them. The total amount of
energy in the cell is limited, so all tumor cell functions will necessarily
slow down. This will lead to retarded growth in tumors, the persistence of
high ITH, and longer survival periods for patients.

\section*{Supporting Information - Appendix A}
In the 2-type population game the expected payoff of $A$-cells at the ESS
increases with $x_{A}$ for $x_{A}<\bar{x}_{A}$ and decreases for $x_{A}>\bar{%
	x}_{A}$.

\begin{proof}
	From (2), compute:  
	\begin{eqnarray*}
		\frac{d}{dx_{A}}\left[ \frac{v}{2}\left(1-\frac{v}{c}\right)+\frac{v^{2}}{2c}\tfrac{
			2n(1-x_{A})-1}{n-nx_{A}-1}\right] &=&\frac{ v^{2}n}{2c\left(
			n-nx_{A}-1\right) ^{2}}>0 \\
		\frac{d}{dx_{A}}\left[ \frac{v}{2}\left(1-\frac{v}{c}\right)+\frac{v}{2c}\frac{
			v(n-1)+cn(1-x_{A})}{n-1}\right] &=&-\frac{vn}{2(n-1)}<0 \\
		\frac{d}{dx_{A}}\left[ \frac{v}{2}\left(1-\frac{v}{c}\right)+\frac{c-v}{c}\frac{
			vn(1-x_{A})}{nx_{A}-1}\right] &=&-\frac{c-v}{c} \frac{vn(n-1)}{\left(
			nx_{A}-1\right) ^{2}}<0.
	\end{eqnarray*}
	Therefore, we have $\frac{d}{dx_{A}}\left[ U_{A}^{\ast \ast }(x_{A})\right] >0$ if
	$x_{A}<\bar{x}_{A}$ and $\frac{d}{dx_{A}}\left[  U_{A}^{\ast \ast }(x_{A})%
	\right] <0$ if $x_{A}>\bar{x}_{A}$.
\end{proof}


\section*{Supporting Information - Appendix B}
\renewcommand{\theequation}{B.\arabic{equation}}
\setcounter{equation}{0}
In a 3-type population game strategy $\left( 0,\frac{(n-1)v/c-nx_{E}}{%
	n-nx_{A}-nx_{E}-1},1\right) $ is an ESS if $x_{E}<\bar{x}_E$ and $x_{A}<\bar{x%
}_A$.

\begin{proof}
	First, we check that strategy $\left( 0,\frac{(n-1)v/c-nx_{E}}{%
		n-nx_{A}-nx_{E}-1},1\right) $ is a best response to itself. Substituting  
	(1) into (3) and then substituting (3) into (4), the expected payoff of a generic cell can be written: 
	\begin{equation}
		U=f(\beta _{A},\beta _{B},\beta _{E})+\sum_{I}f_{I}(\beta _{A},\beta
		_{B},\beta _{E})\alpha _{I},\ I=\left\lbrace A,B,E\right\rbrace
		\label{U3F}
	\end{equation}
	\begin{eqnarray*}
		\text{where }f(\beta _{A},\beta _{B},\beta _{E}) &=&\frac{v}{2}\left[ 1-\sum_I x_{I}\beta	_{I}\right] \nonumber\\
		f_{I}(\beta _{A},\beta _{B},\beta _{E}) &=&\frac{c}{2}\frac{nx_I}{n-1}\left[ \frac{v}{%
			c}\left( 1-\frac{1}{n}\right) -\sum_{I}x_{I}\beta _{I}+\frac{\beta_I}{n}\right],\ I=\left\lbrace A,B,E\right\rbrace\nonumber
	\end{eqnarray*}
	
	The optimal choice of a cell given that its
	opponent plays $(\beta _{A},\beta _{B},\beta _{E})$ 
	is to play $\alpha _{I}=1$ whenever $f_{I}(\beta _{A},\beta _{B},\beta
	_{E})>0$, $\alpha _{I}=0$ whenever $f_{I}(\beta _{A},\beta _{B},\beta
	_{E})<0 $ and $\alpha _{I}\in[0,1]$ whenever $f_{I}(\beta _{A},\beta _{B},\beta
	_{E})=0$ where $I=A,B,E$. If its opponent plays $\left( 0,\beta _{B},1\right) $ the optimal choice of
	a cell is $\alpha _{A}=0$, $\alpha _{B}=\beta _{B}$ (with $0\leq\alpha_{B}\leq 1$)
	and $\alpha _{E}=1$ as long as $f_{A}\left( 0,\beta _{B},1\right) <0$, $%
	f_{B}\left( 0,\beta _{B},1\right) =0$ and $f_{E}\left( 0,\beta _{B},1\right)
	>0$. 
	
	Equality $f_{B}\left( 0,\beta _{B},1\right) =0$ is satisfied if $\frac{v}{%
		c}\left( 1-\frac{1}{n}\right) -(1-x_A-x_E)\beta_B-x_E+\frac{\beta_B}{n}=0$ 
	or equivalently $\frac{v}{c}\left(n-1\right)-(n-nx_A-nx_E)\beta_B-nx_E+\beta_B=0$, leading to $\beta
	_{B}=\frac{(n-1)v/c-nx_{E}}{n-nx_{A}-nx_{E}-1}$. 
	
	Condition $\beta _{B}>0$
	gives $(n-1)v/c-nx_{E}$ $>0$ or equivalently $x_{E}<\bar{x}_E$ (assuming
	there is strictly more than one cell of each type, i.e., $%
	nx_{B}=n-nx_{A}-nx_{E}>1$). 
	
	Condition $\beta _{B}<1$ gives $%
	(n-1)v/c-nx_{E}<n-nx_{A}-nx_{E}-1$, or $x_{A}<\bar{x}_A$. It remains to
	verify that $f_{A}\left( 0,\frac{(n-1)v/c-nx_{E}}{n-nx_{A}-nx_{E}-1}%
	,1\right) <0$ and $f_{E}\left( 0,\frac{(n-1)v/c-nx_{E}}{n-nx_{A}-nx_{E}-1}%
	,1\right) >0$ hold at $\left( 0,\frac{(n-1)v/c-nx_{E}}{n-nx_{A}-nx_{E}-1}%
	,1\right) $, which with simple algebraic manipulations can be proved that is
	true.
	
	Second, it remains to verify that strategy $\left( 0,\frac{(n-1)v/c-nx_{E}}{%
		n-nx_{A}-nx_{E}-1},1\right) $ which is a best response to $\left( 0,\frac{%
		(n-1)v/c-nx_{E}}{n-nx_{A}-nx_{E}-1},1\right) $ cannot be invaded by a mutant
	strategy. 
	
	Let the opponent play $(\gamma _{A},\gamma _{B},\gamma _{E})$, as a
	best response to $\left( 0,\frac{(n-1)v/c-nx_{E}}{n-nx_{A}-nx_{E}-1}%
	,1\right) $. We have to check that the expected payoff obtained when playing $\left( 0,%
	\frac{(n-1)v/c-nx_{E}}{n-nx_{A}-nx_{E}-1},1\right) $ is larger than when it
	is playing $(\gamma _{A},\gamma _{B},\gamma _{E})$. As checked above $%
	f_{A}\left( 0,\frac{(n-1)v/c-nx_{E}}{n-nx_{A}-nx_{E}-1},1\right) <0$, $%
	f_{B}\left( 0,\frac{(n-1)v/c-nx_{E}}{n-nx_{A}-nx_{E}-1},1\right) =0$ and $%
	f_{E}\left( 0,\frac{(n-1)v/c-nx_{E}}{n-nx_{A}-nx_{E}-1},1\right) >0$. Thus,
	we have $\gamma _{A}=0$, $0<\gamma _{B}<1$, and $\gamma _{E}=1$. Then, from (%
	\ref{U3F}) we write the difference of expected payoffs when playing $\left( 0,%
	\frac{(n-1)v/c-nx_{E}}{n-nx_{A}-nx_{E}-1},1\right) $ and $\left( 0,\gamma
	_{B},1\right)$ whenever the opponent plays $\left( 0,\gamma _{B},1\right)$:
	\begin{eqnarray*}
		&&f\left( 0,\gamma _{B},1\right) +f_{A}\left( 0,\gamma _{B},1\right) \cdot
		0+f_{B}\left( 0,\gamma _{B},1\right) \text{ }\frac{(n-1)v/c-nx_{E}}{%
			n-nx_{A}-nx_{E}-1}+f_{E}\left( 0,\gamma _{B},1\right) \cdot 1 \\
		&&-f\left( 0,\gamma _{B},1\right) -f_{A}\left( 0,\gamma _{B},1\right) \cdot
		0-f_{B}\left( 0,\gamma _{B},1\right) \gamma _{B}-f_{E}\left( 0,\gamma
		_{B},1\right) \cdot 1 \\
		&=&f_{B}\left( 0,\gamma _{B},1\right) \left( \frac{(n-1)v/c-nx_{E}}{%
			n-nx_{A}-nx_{E}-1}-\gamma _{B}\right) \\
		&=&\frac{c}{2}(1-x_{A}-x_{E})\left( \frac{v}{c}-\frac{nx_{E}}{n-1}-\frac{%
			n-nx_{A}-nx_{E}-1}{n-1}\gamma _{B}\right) \left( \frac{(n-1)v/c-nx_{E}}{%
			n-nx_{A}-nx_{E}-1}-\gamma _{B}\right) \\
		&=&\frac{c}{2}(1-x_{A}-x_{E})\frac{n-nx_{A}-nx_{E}-1}{n-1}\left( \frac{%
			(n-1)v/c-nx_{E}}{n-nx_{A}-nx_{E}-1}-\gamma _{B}\right) ^{2} \\
		&>&0\text{ if }\gamma _{B}\neq \frac{(n-1)v/c-nx_{E}}{n-nx_{A}-nx_{E}-1}
	\end{eqnarray*}%
	Thus, the expected payoff is strictly larger playing $\left( 0,\frac{(n-1)v/c-nx_{E}}{%
		n-nx_{A}-nx_{E}-1},1\right) $ than $\left( 0,\gamma _{B},1\right) $ and
	strategy $\left( 0,\frac{(n-1)v/c-nx_{E}}{n-nx_{A}-nx_{E}-1},1\right) $ is
	an ESS.
\end{proof}


\section*{Supporting Information - Appendix C}
In a 3-type population game, the aggression suffered by $B$-cells at the
ESS increases with the proportion of $A$-cells, $x_{A}$; while it decreases
with the proportion of $E$-cells, $x_{E}$.

\begin{proof}
	From (5), compute:  
	\begin{equation*}
		\frac{\partial }{\partial x_{A}}\left[ \frac{(n-1)v/c-nx_{E}}{
			n-nx_{A}-nx_{E}-1}\right] =\frac{n\left( (n-1)v/c-nx_{E}\right) }{\left(
			n-nx_{A}-nx_{E}-1\right) ^{2}}>0
	\end{equation*}
	given that $x_{E}<(1-\frac{1}{n})\frac{v}{c}=\bar{x}_E$ at the ESS. 
	\begin{eqnarray*}
		\frac{\partial }{\partial x_{E}}\left[ \frac{(n-1)v/c-nx_{E}}{
			n-nx_{A}-nx_{E}-1}\right] &=&\frac{-n\left( n-nx_{A}-1\right) +n\left(
			(n-1)v/c\right) }{\left( n-nx_{A}-nx_{E}-1\right) ^{2}}<0
	\end{eqnarray*}
	given that $x_{A}<(1-\frac{1}{n})(1-\frac{v}{c})=\bar{x}_A$ at the ESS.
	
	Therefore, for $x_{A}<\bar{x}_A$ and $x_{E}<\bar{x}_E$ at the ESS, we have $\frac{\partial\alpha_B^*}{\partial x_{A}}>0$ and $\frac{\partial\alpha_B^*}{\partial x_{E}}<0$. 
\end{proof}


\section*{Supporting Information - Appendix D}
In the 3-type population game the expected payoff of each $A$-cell at the ESS
increases with $x_{A}$ and decreases with $x_{E}$.

\begin{proof}
	From (6), compute:  
	\begin{eqnarray*}
		\frac{\partial }{\partial x_{A}}\left[ U_{A}^{\ast \ast \ast }(x_{A},x_{E}) %
		\right] &=&\frac{\partial }{\partial x_{A}}\left[ \frac{v}{2}\left(1-\frac{v}{c}\right)+ \frac{v^{2}}{2c}\frac{2n-2nx_{A}-2nx_{E}-1}{n-nx_{A}-nx_{E}-1}-\frac{v}{2}
		\frac{nx_{E}}{\left( n-nx_{A}-nx_{E}-1\right) \left( n-1\right) }\right] \\
		&=&\frac{v}{2}\left[ \frac{v}{c}\frac{(n-1)}{n}-x_{E}\right] \frac{n}{\left(
			n-nx_{A}-nx_{E}-1\right) ^{2}}\frac{n}{n-1}>0.
	\end{eqnarray*}
	given that $x_{E}<(1-\frac{1}{n})\frac{v}{c}=\bar{x}_E$ at the ESS.
	\begin{eqnarray*}
		\frac{\partial }{\partial x_{E}}\left[ U_{A}^{\ast \ast \ast }(x_{A},x_{E}) %
		\right] &=&\frac{\partial }{\partial x_{E}}\left[ \frac{v}{2}\left(1-\frac{v}{c}\right)+ \frac{v^{2}}{2c}\frac{2n-2nx_{A}-2nx_{E}-1}{n-nx_{A}-nx_{E}-1}-\frac{v}{2}
		\frac{nx_{E}}{\left( n-nx_{A}-nx_{E}-1\right) \left( n-1\right) }\right] \\
		&=&\frac{v}{2}\left[ \frac{v}{c}\frac{(n-1)}{n}-\frac{(n-1)}{n}+x_{A}\right] \frac{n}{
			\left( n-nx_{A}-nx_{E}-1\right) ^{2}}\frac{n}{n-1} \\
		&=&\frac{v}{2}\left[ x_{A}-\left(1-\frac{v}{c}\right)\left(1-\frac{1}{n}\right)\right] \frac{n}{
			\left( n-nx_{A}-nx_{E}-1\right) ^{2}}\frac{n}{n-1}<0.
	\end{eqnarray*}
	given that $x_{A}<(1-\frac{1}{n})(1-\frac{v}{c})=\bar{x}_A$ at the ESS.
	
	Therefore, we have $\frac{\partial }{\partial x_{A}}\left[ U_{A}^{\ast \ast \ast }(x_{A},x_{E})%
	\right] >0$ and $\frac{\partial }{\partial x_{E}}\left[ U_{A}^{\ast \ast \ast }(x_{A},x_{E}) %
	\right] <0$.
\end{proof}

\section*{Supporting Information - Appendix E}
For $x_{A}<\bar{x}_A$ and $x_{E}<\bar{x}_E$ we have $U_{A}^{**}(x_{A})>U_{A}^{***}(x_{A},x_{E})$.

\begin{proof}
	From (2) and (6), compute:  
	\begin{equation*}
		U_{A}^{\ast \ast }(x_{A})-U_{A}^{\ast \ast \ast }(x_{A},x_{E}) =\frac{v^{2}}{2c} 
		\left( \frac{2n-2nx_{A}-1}{n-nx_{A}-1}-\frac{2n-2nx_{A}-2nx_{E}-1}{n-nx_{A}-nx_{E}-1}\right) 
		+\frac{v}{2}\frac{nx_{E}}{\left( n-nx_{A}-nx_{E}-1\right) \left( n-1\right) }.
	\end{equation*}
	First, note that:  
	\begin{eqnarray*}
		\frac{v^{2}}{2c}\left( \frac{2n-2nx_{A}-1}{n-nx_{A}-1}-\frac{2n-2nx_{A}-2nx_{E}-1}{n-nx_{A}-nx_{E}-1}\right) &=&\frac{v^{2}}{2c}\frac{
			(2n-2nx_{A}-1)(-nx_{E})+2nx_{E}(n-nx_{A}-1)}{(n-nx_{A}-1)(n-nx_{A}-nx_{E}-1)}
		\\
		&=&\frac{v^{2}}{2c}\frac{-nx_{E}}{(n-nx_{A}-1)(n-nx_{A}-nx_{E}-1)}.
	\end{eqnarray*}
	This gives:  
	\begin{eqnarray*}
		U_{A}^{\ast \ast }(x_{A},x_{E})-U_{A}^{\ast \ast \ast }(x_{A},x_{E}) &=& 
		\frac{v^{2}}{2c}\frac{-nx_{E}}{(n-nx_{A}-1)(n-nx_{A}-nx_{E}-1)}+\frac{v}{2} 
		\frac{nx_{E}}{\left( n-nx_{A}-nx_{E}-1\right) \left( n-1\right) } \\
		&=&\frac{v}{2}\frac{nx_{E}}{\left( n-nx_{A}-nx_{E}-1\right) \left(
			n-1\right) (n-nx_{A}-1)}\left[ -\frac{v}{c}(n-1)+(n-nx_{A}-1)\right] \\
		&=&\frac{v}{2}\frac{nx_{E}}{\left( n-nx_{A}-nx_{E}-1\right) \left(
			n-1\right) (n-nx_{A}-1)}\left[(n-1)\left(1-\frac{v}{c}\right)-nx_{A}\right]
		\\
		&>&0\text{ given that }x_{A}<\left(1-\frac{1}{n}\right)\left(1-\frac{v}{c}\right)=\bar{x}_A.
	\end{eqnarray*}
	Thus, for $x_{A}<\bar{x}_A$ and $x_{E}<\bar{x}_E$, we have $U_{A}^{**}(x_{A})>U_{A}^{***}(x_{A},x_{E})$. 
\end{proof}

\end{document}